\documentclass[10pt,twocolumn,journal]{IEEEtran}
\usepackage{latexsym}
\usepackage{cite}
\usepackage{amsmath,amssymb,amsfonts,bm,amsthm}

\usepackage{cases}
\usepackage{cite}
\usepackage{algorithm}
\usepackage{algorithmic}
\usepackage{graphicx}
\usepackage{textcomp}
\usepackage{xcolor}
\usepackage{multirow}
\usepackage{array}
\usepackage{subfig}
\usepackage{caption}
\usepackage{lipsum}
\newtheorem{lemma}{Lemma}

\usepackage[T1]{fontenc}
\usepackage{colortbl}

\begin{document}
\title{\LARGE Latency Minimization in Intelligent Reflecting Surface Assisted D2D Offloading Systems}


\author{Yanzhen Liu, Qiyu Hu,  Yunlong Cai, \IEEEmembership{Senior Member,~IEEE,} and Markku Juntti, \IEEEmembership{Fellow,~IEEE}
	\thanks{
		Y. Liu, Q. Hu, and Y. Cai are with the College of Information Science and Electronic Engineering, Zhejiang University, Hangzhou 310027, China (e-mail: yanzliu@zju.edu.cn; qiyhu@zju.edu.cn; ylcai@zju.edu.cn). M. Juntti is with the Centre for Wireless Communications, University of Oulu,  Oulu 90014, Finland (e-mail: markku.juntti@oulu.fi).
	}
}

\maketitle
\vspace{-3.3em}
\begin{abstract}

In this letter, we investigate an intelligent reflecting surface (IRS) aided device-to-device (D2D) offloading system, where an IRS is employed to assist in computation offloading from a group of users with intensive tasks to another group of idle users. We propose a new two-timescale joint passive beamforming and resource allocation algorithm based on stochastic successive convex approximation to minimize the system latency while cutting down the heavy overhead in exchange of channel state information (CSI). Specifically, the high-dimensional passive beamforming vector at the IRS is updated in a frame-based manner based on the channel statistics, where each frame consists of a number of time slots, while the offloading ratio and user matching strategy are optimized relied on the low-dimensional real-time effective channel coefficients in each time slot. The convergence property and the computational complexity of the proposed algorithm are also examined. Simulation results show that our proposed algorithm significantly outperforms the conventional benchmarks.
\end{abstract}


\IEEEpeerreviewmaketitle
\vspace{-1.2em}
\section{Introduction}

Recently, mobile edge computing (MEC) is considered as an effective technique to reduce the computation latency of  intensive-task applications with the aid of both local computing on the devices and edge computing \cite{Survey}. However, the computing resources of the edge server are limited and its heavy computing load needs to be alleviated. Hence, the device-to-device (D2D) communication has been employed for reducing the burden of the edge server through user collaborative offloading \cite{Quality,Stochastic,Joint,D2DMEC}.
However, the potential of D2D computation offloading has not been fully exploited since the offloading link between the users is far from perfect \cite{Quality}. Specifically, the D2D users that are far from each other typically suffer from a low offloading success rate due to the limited transmit power. Moreover, the communication links used for offloading tasks are very likely to be blocked by the obstructions, especially in the indoor environment. Therefore, it is necessary to improve the performance of D2D offloading systems from a communication perspective.

Fortunately, the propagation-induced impairments can be mitigated by the intelligent reflecting surface (IRS), which has been envisioned as an innovative hardware-efficient technology for the beyond fifth-generation (B5G) wireless system \cite{IRSsurvey1,IRSsurvey2}. IRS can modify the signal propagation by dynamically adjusting its reflection coefficients such that the desired and interfering signals can be enhanced and suppressed, respectively. As a result, by smartly coordinating these reflecting elements, IRS is able to create a favorable signal propagation environment to improve the wireless communication coverage, throughput, and energy efficiency substantially \cite{IRS,Latency,IRSEnhanced}. The authors of \cite{IRS} aimed at jointly optimizing the active beamforming at the access point (AP) and the passive beamforming at the IRS to minimize the power consumption. In \cite{Latency}, a block coordinate descent (BCD) technique has been developed to minimize the latency for the IRS-aided MEC system. Further, the authors of \cite{IRSEnhanced} have developed a mixed-timescale algorithm via exploiting the channel statistics to reduce the system overhead in an IRS enhanced system.



To the best of our knowledge, although the IRS-assisted computation offloading to the edge server has been investigated in \cite{Latency}, the IRS-aided D2D offloading system has not been well studied. In this work, we investigate an IRS-aided D2D offloading system, where an IRS is employed to assist in computation offloading from a group of users with intensive tasks to another group of idle users. We propose a new two-timescale joint passive beamforming and resource allocation algorithm based on stochastic successive convex approximation (SSCA) to minimize the system latency while cutting down the heavy overhead for CSI feedback. Specifically, the high-dimensional passive beamforming vector at the IRS is updated in a frame-based manner based on the channel statistics, where each frame consists of a number of time slots, while the offloading ratio and user matching strategy are optimized relied on the low-dimensional real-time effective channel coefficients in each time slot. The convergence property and the computational complexity of the proposed algorithm are also examined. Simulation results show that our proposed algorithm significantly outperforms the conventional benchmarks.

\emph{Notations:} Scalars, vectors and matrices are denoted by lower case, boldface lower case and boldface upper case letters, respectively. For a  matrix $\mathbf{A}$, ${{\mathbf{A}}^T}$, $\textrm{conj}(\mathbf{A})$, and ${{\mathbf{A}}^H}$ denote its transpose, conjugate and conjugate transpose, respectively. The imaginary unit is denoted by $j$ and $| \cdot |$ denotes the absolute value of a complex scalar. $\circ$ is the Hadamard product and $\angle$ computes the phase of a complex scalar or a complex vector element-wise.
\vspace{-1.2em}
\section{System Model and Problem Formulation}

\begin{figure}[!t]
\centering
\scalebox{0.78}{\includegraphics{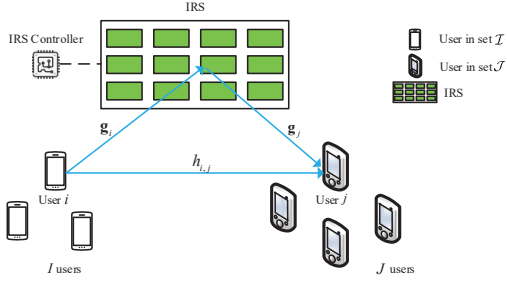}}
\caption{IRS-aided D2D offloading system.}\label{system}
\end{figure}

We consider an IRS-assisted D2D offloading system consisting of $I$ users which have intensive tasks to be processed, indexed by set $\mathcal{I}\triangleq\{1, 2, \ldots, I\}$, $J$ ($J\geq I$) idle users which are able to provide computing services, indexed by $\mathcal{J}\triangleq \{I+1,I+2, \ldots, I+J\}$, and an IRS equipped with a controller and $M$ reflecting elements, as illustrated in Fig. \ref{system}. We assume partial offloading model \cite{Survey}, thus the users in $\mathcal{I}$ can divide their tasks into two parts. One portion is computed by the local CPU and the other portion is offloaded to a scheduled user in $\mathcal{J}$ and processed by the CPU of the helper.

\vspace{-1.1em}
\subsection{Computation Model}
By following the computation model in \cite{D2DMEC}, we use a tuple $\{L_i$, $C_i\}$ to characterize the task of user $i$ in set $\mathcal{I}$, where $L_i$ (in bits) denotes the size of the task and $C_i$ (in CPU cycles/bit) denotes the CPU cycles required to compute 1-bit data of the task at user $i$. Moreover, we let $f_i$ and $f_j$ (in CPU cycles/s) denote the computing resource of user $i$ in set $\mathcal{I}$ and user $j$ in set $\mathcal{J}$, respectively. As mentioned above, we adopt the partial offloading strategy, i.e., user $i$ in set $\mathcal{I}$ offloads $\rho_{i,j}L_i$ bits to a matched D2D user $j$ in set $\mathcal{J}$, where $0\leq \rho_{i,j} \leq1$ denotes the offloading ratio, and the remaining $(1-\rho_{i,j})L_i$ bits are computed locally. Hence, the delay of the local computing can be expressed as
\begin{equation}
	T_{i,j}^L=\frac{(1-\rho_{i,j})C_iL_i}{f_i}. \label{til}
\end{equation}
By neglecting the feedback delay for the length of computation results is generally short \cite{Stochastic,D2DMEC}, the D2D offloading delay can be expressed as the summation of the transmission delay and computation delay:
\begin{equation}
	T_{i,j}^D=\frac{\rho_{i,j}L_i}{r_{i,j}}+\frac{\rho_{i,j}C_iL_i}{f_j}, \label{tid}
\end{equation}
where $r_{i,j}$ is the communication rate between user $i$ and user $j$. Then, we express the total delay for the task processing of user $i$ as
\begin{equation}
	T_{i,j}=\text{max}\{T_{i,j}^L,T_{i,j}^D\}. \label{ti}
\end{equation}

Without loss of generality, we assume that the users in $\mathcal{I}$ only can offload tasks to one user in set $\mathcal{J}$ as in \cite{D2DMEC}. Moreover, we define a binary decision variable $u_{i,j}$ to indicate the status of the D2D link established between the two groups of users, i.e., $u_{i,j}=1$ when user $i$ establishes a D2D link with user $j$ and $u_{i,j}=0$ otherwise.
\vspace{-1.1em}
\subsection{Communication Model}
We adopt the orthogonal frequency-division multiple access (OFDMA) and each D2D link is allocated with one sub-channel. For the sake of exposition, we assume that the users are all equipped with a single antenna. For a given D2D pair of user $i \in \mathcal{I}$ and user $j \in \mathcal{J}$, we let $h_{i,j} \in \mathbb{C}$ denote the channel coefficient between user $i$ and user $j$, $\bm{g}_{i} \in \mathbb{C}^{M\times 1}$ denote the channel vector between user $i$ and the IRS, and $\bm{g}_{j} \in \mathbb{C}^{M\times 1}$ denote the channel vector between user $j$ and the IRS. Let $s_i\sim\mathcal{CN}(0,1)$ denote the transmit symbol of user $i$. Then, the received signal at user $j$ is given as
\begin{equation}
	y_j=\sqrt{p_i}(h_{i,j}+\bm{g}_{j}^H\bm{\Phi}\bm{g}_{i})s_i+n_j,
\end{equation}
where $p_i$ denotes the transmit power of user $i$ and $n_j \sim\mathcal{CN}(0,\sigma_j^2)$ denotes the additive white Gaussian noise with zero mean and variance $\sigma^2_j$. $\bm{\Phi} \in \mathbb{C}^{M\times M}$ denotes the passive beamforming matrix at the IRS, which is a diagonal matrix due to no signal processing over its passive reflecting elements. We define the passive beamforming vector $\bm{\phi}\in\mathbb{C}^{M\times 1}\triangleq \mathrm{diag}\{\mathbf{\Phi}\}$. Then, we obtain the maximum achievable transmission rate between user $i$ and user $j$ as
\begin{equation}
	r_{i,j}=B\log(1+\frac{p_i|h_{i,j}+\bm{g}_{i,j}^H\bm{\phi}|^2}{\sigma_j^2}), \label{rijr} \vspace{-2.0mm}
\end{equation}
where $B$ is the bandwidth of the sub-channel allocated to each D2D link and $\bm{g}_{i,j} \triangleq \textrm{conj}(\bm{g}_{i}) \circ \bm{g}_{j}$.

\vspace{-1.1em}
\subsection{Timescale Model}
\begin{figure}[t]
	\begin{centering}
		\includegraphics[width=3.2in]{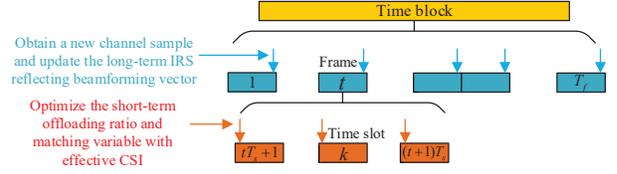}
	\end{centering}
	\caption{An illustration of two-timescale frame structure.}
	\label{TwoTime}
\end{figure}
\begin{figure*}[!t]
	\begin{equation}
		\begin{split}
			\frac{\partial g(\boldsymbol{\theta},\bm{x}^\star)}{\partial \boldsymbol{\theta}} &= \sum_{i=1}^{I} -\frac{w_iL_iC_i^2f_{\pi_i}^2}{(f_if_{\pi_i}+C_ir_{i,\pi_i}(f_i+f_{\pi_i}))^2}\frac{p_iB}{\sigma_{\pi_i}^2+p_i|h_{i,\pi_i}+\bm{g}_{i,\pi_i}^H\bm{\phi}|^2}  \\
			&\times(\text{conj}(h_{i,\pi_i}\bm{g}_{i,\pi_i}+\bm{g}_{i,\pi_i}\bm{g}_{i,\pi_i}^H\bm{\phi})\circ je^{j\boldsymbol{\theta}}-(h_{i,\pi_i}\bm{g}_{i,\pi_i}+\bm{g}_{i,\pi_i}\bm{g}_{i,\pi_i}^H\bm{\phi})\circ je^{j\boldsymbol{\theta}}). \label{gradient}
		\end{split} \tag{15}
	\end{equation}
\end{figure*}
The joint optimization of the IRS passive beamforming and task allocation for each CSI realization is not practical for
implementation, since it entails a huge amount of overhead in exchange of high dimensional real-time CSI. To address this issue, we investigate a two-timescale scheme that considers both the CSI statistics and low-dimensional effective instantaneous CSI. As illustrated in Fig. \ref{TwoTime},
we focus on a sufficiently large time block, during which the channel statistics are supposed to be constant. It consists of $T_f$ frames, each of which is further divided into $T_s$ time slots, and we assume that the CSI remains invariant within each time slot. Based on this assumption, we define the following concepts of timescales:
\begin{itemize}
	\item Long-timescale: The channel statistics (distribution) are assumed constant over each time block consisting of  $T_f$ frames.
	\item Short-timescale: The channel gains are assumed invariant during each time slot.
\end{itemize}

In each frame, the devices obtain a CSI sample $(\bm{g}_{i,j},h_{i,j})$. Then, in each time slot, they can acquire the real-time effective channel coefficient $\tilde{h}_{i,j}\triangleq h_{i,j}+\bm{g}_{i,j}^H\bm{\phi}$. The long-term IRS passive beamforming vector is updated at the end of each frame based on a CSI sample and the short-term offloading ratio and user matching strategy are optimized in each time slot by using the low-dimensional effective CSI.

\vspace{-1.1em}
\subsection{Problem Formulation}
We formulate the latency minimization problem as
\begin{subequations}
	\begin{align}
		(\mathcal{P})\min_{u_{i,j},\bm{\phi} \atop 0\leq \rho_{i,j} \leq 1} \quad&\sum_{i=1}^{I}\sum_{j=I+1}^{I+J} \mathbb{E}_{\mathbf{H}}\{u_{i,j}w_iT_{i,j}\}  \\
		\text{s.t.} \quad
		&u_{i,j}\in\{0,1\},\forall i,j, \label{u_cst1}\\
		&\sum_{i=1}^{I}u_{i,j}\leq 1,\forall j, \quad \sum_{j=I+1}^{I+J}u_{i,j}\leq 1,\forall i, \label{u_cst2}\\
		&|\bm{\phi}(m)|=1,\forall m, \label{phi_cst}	
		\end{align}
\end{subequations}
where the fixed weight $w_i$ is used to represent the priority of the tasks and $\mathbf{H}\triangleq \{(\bm{g}_{i,j},h_{i,j}),\forall i,j\}$ denotes the channel set. Constraints (\ref{u_cst1}) and (\ref{u_cst2}) guarantee that each user establishes only one D2D link, and constraint (\ref{phi_cst}) denotes the unit modulus constraint on the elements of the IRS passive beamforming vector.
\vspace{-0.8em}
\section{Proposed Two-timescale  Algorithm}
As we can see, $\mathcal{P}$ is a mixed integer non-linear problem (MINLP) with non-convex stochastic objective function and unit modulus constraints, which is very challenging to solve. In this section, we develop an efficient online SSCA based algorithm to tackle this problem.
\vspace{-1.1em}
\subsection{Short-term Offloading Ratio and User Matching Strategy}
With fixed long-term IRS passive beamforming vector $\bm{\phi}^t$ in frame $t$, for given real-time effective channel coefficients $ \{\tilde{h}^k_{i,j},\forall i,j\}$ in time slot $k$, the short-term optimization problem of the offloading ratio and user matching strategy design yields
\begin{subequations}
	\begin{align}
		(\mathcal{P}1)\min_{u_{i,j},\atop 0\leq\rho_{i,j}\leq 1}  \quad&\sum_{i=1}^{I}\sum_{j=I+1}^{I+J} u_{i,j}w_iT_{i,j}  \\
		\text{s.t.} \quad
		&\eqref{u_cst1},\eqref{u_cst2}.
	\end{align}
\end{subequations}

Note that $\rho_{i,j}$ is not coupled with $u_{i,j}$ in the constraints. Hence we can first optimize the offloading ratio to minimize $T_{i,j}$, which provides the following $I\times J$ parallel subproblems
\begin{equation}
		\min_{0\leq \rho_{i,j} \leq1} \quad \max \{T_{i,j}^L,T_{i,j}^D\} \,\,\,\forall i,j.
\end{equation}
Since both $T_{i,j}^L$ and $T_{i,j}^D$ are linear functions of $\rho_{i,j}$, it is readily seen that the optimal $\rho^\star_{i,j}$ should satisfy $T_{i,j}^L =T_{i,j}^D$, and we obtain
\begin{equation}
	\rho_{i,j}^\star=\frac{C_if_jr_{i,j}}{C_i(f_i+f_j)r_{i,j}+f_if_j}. \label{rhovalue}
\end{equation}
By substituting (\ref{rhovalue}) into (\ref{ti}), we obtain
\begin{equation}
	T_{i,j}^\star=\frac{C_iL_i}{f_i+f_j}+\frac{C_iL_if_j^2}{f_if_j(f_i+f_j)+C_ir_{i,j}(f_i+f_j)^2}. \label{T_ij}
\end{equation}
Then, based on \eqref{T_ij}, problem $\mathcal{P}$1 with respect to $u_{i,j}$ can be viewed as a bipartite graph maximum matching problem with weight $-w_iT^\star_{i,j}$ between any given pair of user $i$ and $j$. The optimal solution $u^\star_{i,j}$ can be efficiently obtained via the celebrated Kuhn-Munkres (KM) algorithm with computational complexity of $\mathcal{O}(I^2J)$ \cite{KM}.

\vspace{-1.0em}
\subsection{Long-term Passive Beamforming Design}
\begin{figure*}[t]
	\begin{minipage}{0.32\linewidth}
		\centering
		\includegraphics[width=2.2in, height=1.8in]{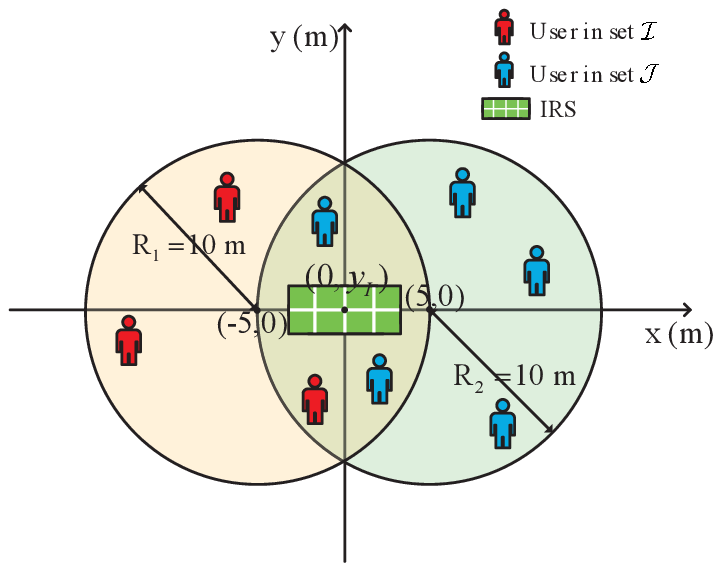}
		\caption{Simulation setup.}
		\label{topview}
	\end{minipage}
	\hfill
	\begin{minipage}{0.32\linewidth}
		\centering
		\includegraphics[width=2.2in, height=1.8in]{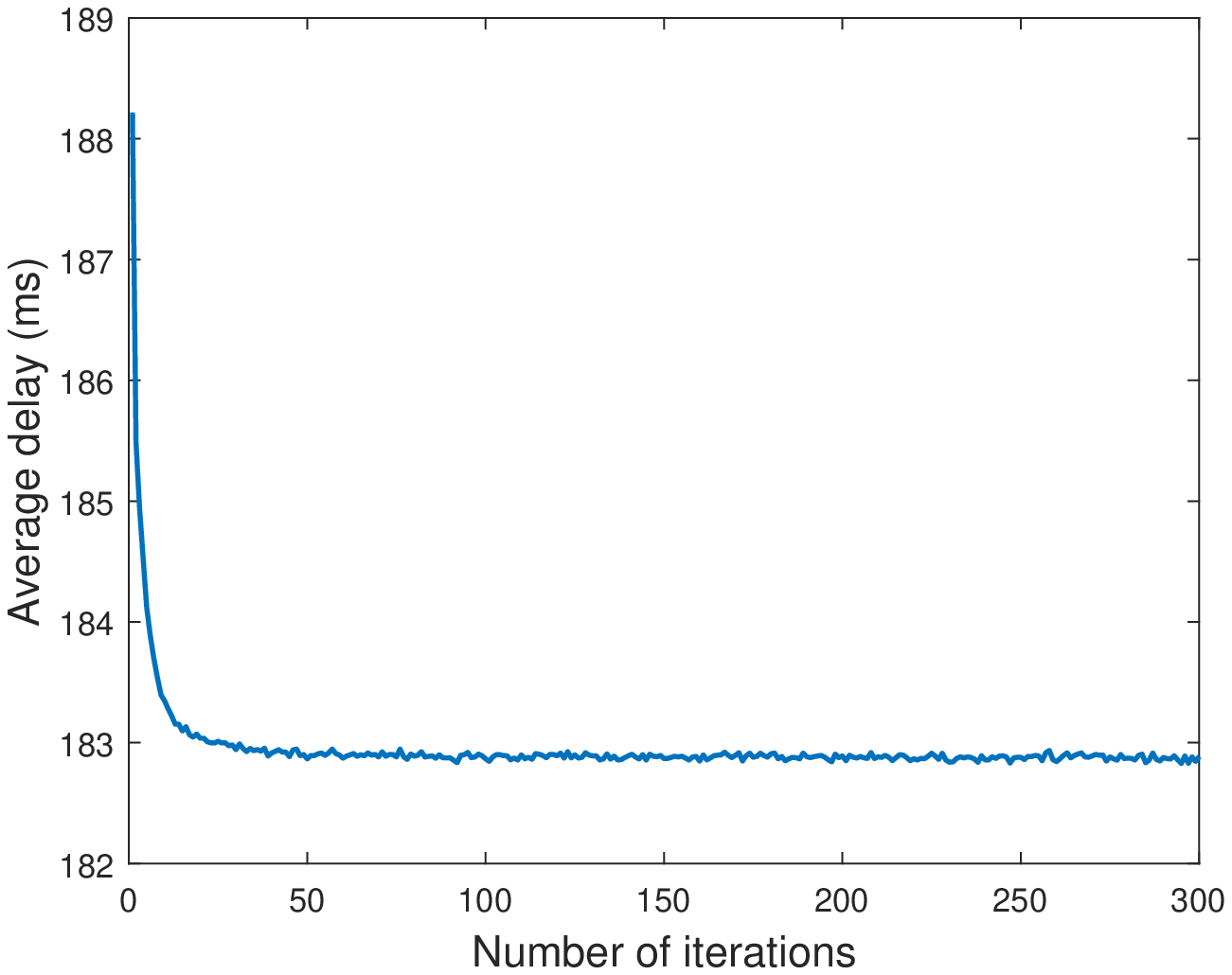}
		\caption{Convergence performance of Algorithm 1.}
		\label{convergence}
	\end{minipage}
	\hfill
	\begin{minipage}{0.32\linewidth}
		\centering
		\includegraphics[width=2.2in, height=1.8in]{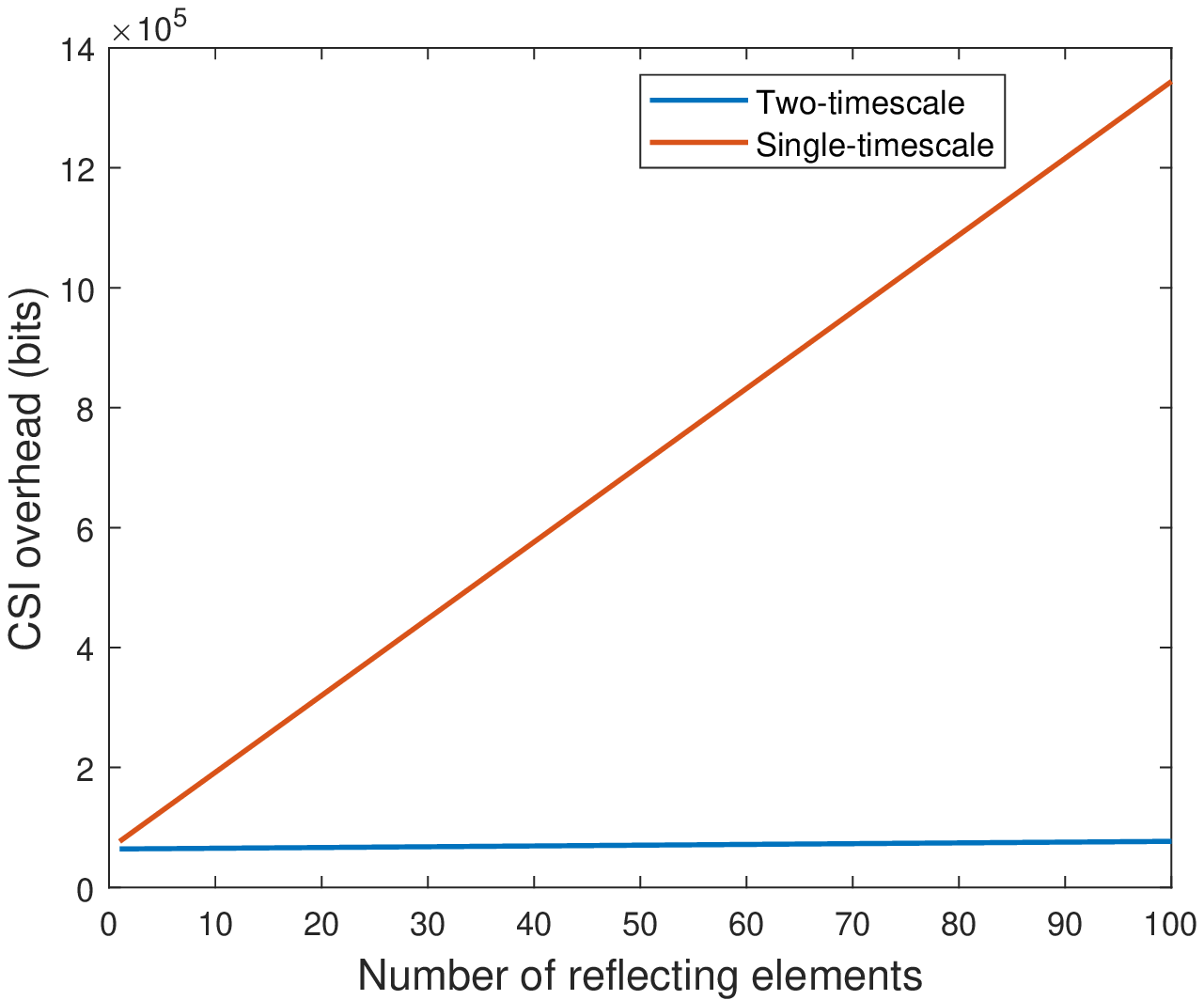}
		\caption{CSI overhead versus the number of reflecting elements.}
		\label{CSIdimension}
	\end{minipage}
\end{figure*}

The long-term optimization problem of the passive beamforming vector $\bm{\phi}$ is given by
\begin{equation}
		(\mathcal{P}2)\,\,\min_{\boldsymbol{\theta}} \quad f(\bm{\theta},\bm{x}^\star) = \mathbb{E}_{\mathbf{H}}\{g(\bm{\theta},\bm{x}^\star)\} \label{long-term-object}  \vspace{-1.5mm}
\end{equation}
where $\boldsymbol{\theta} \triangleq \angle{\bm{\phi}}$, $\bm{x}^\star\triangleq[\rho^\star_{i,j},u^\star_{i,j}]$ denoting the collection of the optimal short-term variables and
\begin{equation}
	g(\bm{\theta},\bm{x}^\star) \triangleq \sum_{i=1}^{I}w_iT^\star_{i,\pi_i}(\boldsymbol{\theta}), \vspace{-1.5mm}
\end{equation}
where we define $\pi_i \triangleq \sum_{j=I+1}^{I+J} u^\star_{i,j}j$ for denoting the user in $\mathcal{J}$ which is matched with user $i$.

Based on the SSCA optimization framework \cite{cssca}, we seek to approximate the original objective function \eqref{long-term-object} by using a quadratic surrogate function. Specifically, at the end of each frame $t$, the channel samples $\{(\bm{g}_{i,\pi_i}^t,h_{i,\pi_i}^t), \forall i\}$ are obtained and the surrogate objective function is updated based on the CSI samples and the short-term variables $\bm{x}^{\star}$ in time slot $(t+1)T_s$ as
\begin{equation}
	\begin{split}
		\bar{f}^t(\boldsymbol{\theta}) =  (\mathbf{f}^t)^T(\boldsymbol{\theta}-\boldsymbol{\theta}^t)+\varpi\|\boldsymbol{\theta}-\boldsymbol{\theta}^t\|^2, \label{surrogate function}
	\end{split}
\end{equation}
where $\boldsymbol{\theta}^t$ is the current value of $\boldsymbol{\theta}$, $\varpi>0$ is a constant, and $\mathbf{f}^t$ denotes the approximation of the partial derivatives $\frac{\partial f}{\partial \boldsymbol{\theta}}$, which is updated based on the following expression
\begin{equation}
	\mathbf{f}^t = (1-\varrho^{t})\mathbf{f}^{t-1}+\varrho^t\frac{\partial g}{\partial \boldsymbol{\theta}}|_{(\boldsymbol{\theta}^t,\bm{x}^{\star})}, \label{surrogate_gradient}
\end{equation}
where $\{\varrho^t\}$ is a sequence to be properly chosen and the expression of $\frac{\partial g}{\partial \boldsymbol{\theta}}$ is given by \eqref{gradient}.

Subsequently, we aim to solve the approximated problem at time frame $t$, which is given by
\addtocounter{equation}{1}
\begin{equation}
	\min_{\boldsymbol{\theta}} \quad \bar{f}^t(\boldsymbol{\theta}). \vspace{-0.5mm}
\end{equation}
This is a convex quadratic problem and the solution can be readily derived as
\begin{equation}
	\bar{\boldsymbol{\theta}}^t = \boldsymbol{\theta}^t-\frac{\mathbf{f}^t}{2\varpi}. \label{solution_to_surrogate}
\end{equation}
Then, the long-term variable is updated as
\begin{equation}
	\boldsymbol{\theta}^{t+1}=(1-\gamma^t)\boldsymbol{\theta}^{t} + \gamma^t\bar{\boldsymbol{\theta}}^t,  \label{theta_update}
\end{equation}
where similarly $\{\gamma^t\}$ denotes a sequence of parameters and the convergence can be guaranteed if we choose $\varrho^t$ and $\gamma^t$ by following the conditions $\lim_{t\rightarrow \infty} \varrho^t = 0, \sum_{t} \varrho^t  = \infty, \sum_{t} (\varrho^t)^2  < \infty,
\lim_{t\rightarrow \infty} \gamma^t = 0, \sum_{t} \gamma^t  = \infty, \sum_{t} (\gamma^t)^2  < \infty$ and $\lim_{t\rightarrow \infty} \frac{\gamma^t}{\varrho^t} = 0$. The convergence proof is given in \textbf{Appendix A}.

The overall two-timescale algorithm is summarized in Algorithm 1\footnote{It is worth noting that the proposed algorithm also applies to the situation of $I> J$ by adding $I-J$ blank users with $f_j=0$ to set $\mathcal{J}$.} and the computational complexity is given by $\mathcal{O}(I^2J+\frac{1}{T_s}IM)$.  In practice, this algorithm can be implemented  on a user in set $\mathcal{J}$, which is referred to as  master user. Specifically, for each time slot, each user $j$ in set $\mathcal{J}$ estimates the effective channel coefficients $\{\tilde{h}_{i,j},\forall i\}$ and feeds them to the master user which performs the proposed algorithm to obtain the short-timescale variables, i.e., the offloading ratios and user matching variables. The offloading ratios are fed back to the users in set $\mathcal{I}$ through signaling channels, and the matched users in set $\mathcal{J}$ are notified to provide services. At the end of each frame, the full channel samples $\{(\bm{g}_{i,\pi_i},h_{i,\pi_i}),\forall i\}$ are estimated at these scheduled users in set $\mathcal{J}$  using the IRS related channel estimation algorithms \cite{IRSCM1,IRSCM2} and then sent to the master user. Finally, the long-timescale  IRS passive beamforming vector is updated based on the proposed algorithm. As a result, we can obtain that the required CSI overhead of the proposed two-timescale algorithm in a frame is $T_sIJ+IM$, while that of the single-timescale algorithm is $T_s(IJ+IM)$, which is significantly reduced.

\begin{algorithm}[t]\caption{Proposed two-timescale algorithm}
	\begin{algorithmic}[1]
		\footnotesize
		\begin{small}
			\STATE Initialize the long-term variable $\boldsymbol{\theta}^0$. Set $f^{-1}=0$, $\mathbf{f}^{-1}=\mathbf{0}$, $t=0$, and $k=0$. Choose proper sequences $\{\varrho^t\}$, $\{\gamma^t\}$ and set a proper value for $\varpi$.
			\REPEAT
			\STATE Obtain the effective CSI $\{\tilde{h}^k_{i,j}, \forall i,j\}$ for time slot $k$.
			\STATE Compute $\rho_{i,j}^\star,\forall i,j$ and $T^\star_{i,j},\forall i,j$ based on \eqref{rhovalue} and \eqref{T_ij}, respectively. Obtain the optimal matching variable $u^\star_{i,j}$ via the KM algorithm.
			\STATE Update $k=k+1$.
			\UNTIL the frame ends, i.e., $k=(t+1)T_s$.			 	
			\STATE Obtain the CSI samples $\{(\bm{g}_{i,\pi_i}^t,h_{i,\pi_i}^t), \forall i\}$. Compute the surrogate function \eqref{surrogate function} based on \eqref{surrogate_gradient}.
			\STATE Obtain the optimal solution via \eqref{solution_to_surrogate} and update $\boldsymbol{\theta}^t$ based on \eqref{theta_update}.
			\STATE Update $t=t+1$ and return to step 2.
		\end{small}
	\end{algorithmic}
\end{algorithm}
\vspace{-1.0em}
\section{Simulation Results}

In this section, we present simulation results to verify the effectiveness of our proposed two-timescale algorithm. A top view of the simulation setup is given in Fig. \ref{topview}. The users in $\mathcal{I}$ and $\mathcal{J}$ are randomly located in two circles with radius of $R_1 = 10\,\text{m}$ and $R_2=10\,\text{m}$, respectively. The coordinates of the centers for these two circles are $(-5\, \text{m},0\, \text{m})$ and $(5\,\text{m},0\,\text{m})$, respectively, and the IRS is located at $(0\,\text{m},y_{\text{I}}=0\,\text{m})$. The heights of the users are all set to $1\,\text{m}$ and the height of the IRS is $3\,\text{m}$. Unless otherwise specified, we consider $I=8$, $J=10$, $M=40$, $T_s =100$, $T_f = 300$, $w_i = 1,\,\forall i$, $f_i = 1\times 10^9$ CPU cycles/s, $\forall i$, $p_i=24$ dBm, $\forall i$, $C_i = 12$ CPU cycles/bit, $\forall i$, $B=2$ MHz, and the noise spectral density as $-174 $ dBm/Hz. Furthermore, the length of the computation tasks $L_i$ is uniformly distributed from $1$ Mbits to $5$ Mbits and the computation resource of user $j$ follows the uniform distribution within $f_j \in [0.5\times 10^9,2.5\times 10^9]$ CPU cycles/s. We adopt the Rician channel with a Rician factor $\beta = 3$ dB~\cite{IRSEnhanced} and the path loss is modeled as $P_{LS} = C_0(\frac{d_{link}}{D_0})^{-\alpha}$, where $C_0$ is the path loss at the reference distance $D_0 = 1\,\text{m}$ and is set to $C_0 = -30$ dB, $d_{link}$ is the link distance, and $\alpha$ is the path loss exponent where we set it for the link between the users as $\alpha_{uu} =3.2$ and the link between the user and IRS as $\alpha_{uI} = 2.2$~\cite{Latency}, respectively.

The performance of the proposed two-timescale algorithm is compared with that of other benchmark approaches. Specifically, we consider the following algorithms:
\begin{itemize}	
	\item Proposed TTS: We employ Algorithm 1 for designing the offloading ratio, the matching strategy, and the IRS passive beamforming vector.
	\item STS: We employ a single-timescale scheme which optimizes the offloading ratio and the matching strategy, as well as the passive beamforming vector at the IRS, based on the real-time high-dimensional full CSI in each time slot.
	\item Max-to-max TTS: We employ a heuristic user matching scheme which repeatedly matches the user in set $\mathcal{I}$ that has the maximum task size $L_i^{max}$ to the user in set $\mathcal{J}$ that has the strongest computation capacity $f_j^{max}$ and removes the matched users from sets $\mathcal{I}$ and $\mathcal{J}$. The offloading ratio and the passive beamforming vector are optimized based on the corresponding steps in Algorithm 1.			
	\item Random IRS: We employ a scheme where the reflection coefficients of the IRS are randomly generated. Then, the offloading ratio is designed based on \eqref{rhovalue} and the matching strategy is optimized based on the KM algorithm.	
	\item No IRS: We do not employ the IRS in the system. The offloading ratio and the matching strategy are designed based on the random IRS method.
\end{itemize}

\begin{figure*}[t]
	\begin{minipage}{0.32\linewidth}
	\centerline{\includegraphics[width=2.3in, height=2.0in]{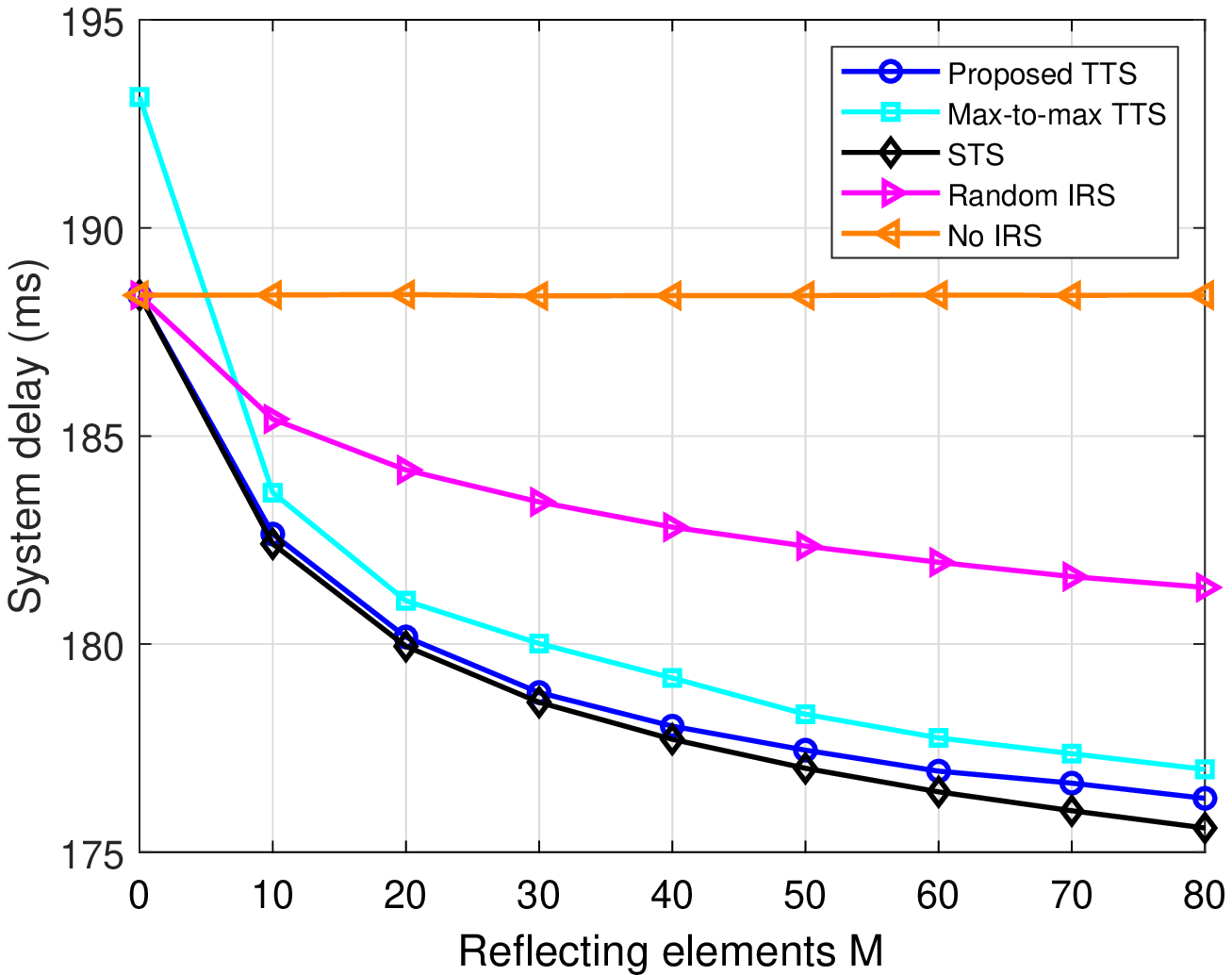}}
\caption{System delay versus the number of reflecting elements $M$.}
\label{performance_reflecting_elements}
	\end{minipage}
	\hfill
	\begin{minipage}{0.32\linewidth}
	\centerline{\includegraphics[width=2.3in, height=2.0in]{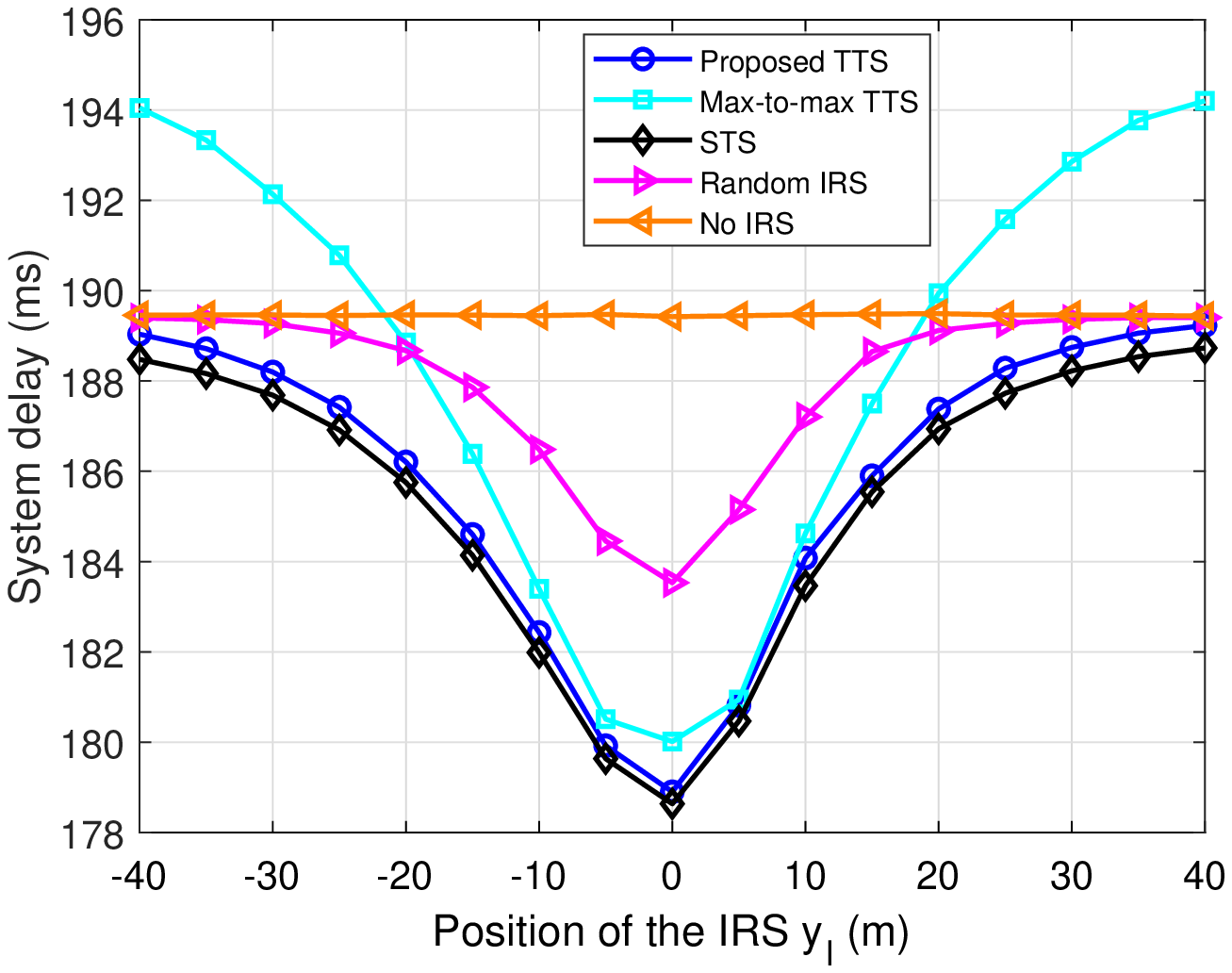}}
\caption{System delay versus the position of the IRS $y_I$.}
\label{performance_IRS_position}
	\end{minipage}
	\hfill
	\begin{minipage}{0.32\linewidth}
	\centerline{\includegraphics[width=2.3in, height=2.0in]{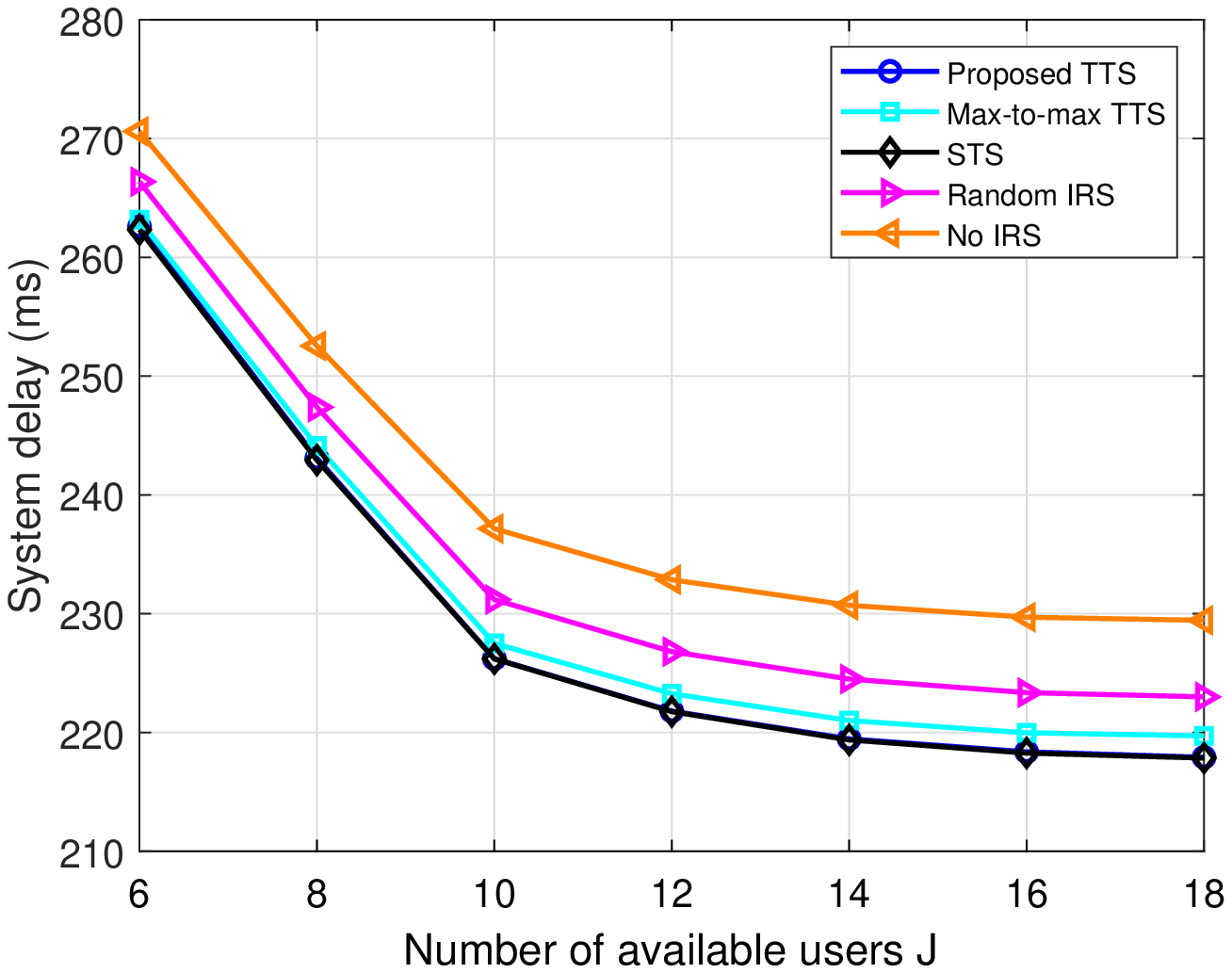}}
\caption{System delay versus the number of available users $J$ ($I=10$).}
\label{performance_user_num}
	\end{minipage}
\end{figure*}

Fig. \ref{convergence} shows the convergence performance of Algorithm 1. As we can see, the average system delay converges quickly within 50 iterations. Fig. \ref{CSIdimension} compares the STS scheme and the proposed TTS scheme in terms of required CSI overhead in a frame, where we assume that the number of quantization bits for each element of the CSI vector equals 8. We can conclude that our proposed TTS algorithm can significantly reduce the CSI overhead compared to the STS scheme. Hence the proposed algorithm is much more suitable for practical design.

Fig. \ref{performance_reflecting_elements} shows the weighted sum system delay of different schemes versus the number of reflecting elements of the IRS. As we can see, the proposed TTS algorithm significantly outperforms the schemes of no IRS, random IRS, and max-to-max TTS. In addition, we observe a small gap between our proposed TTS design and the STS scheme, especially when the number of reflecting elements is small. This is because the line-of-sight (LOS) component dominates the IRS related links, thus rendering valid designs based on the channel statistics.
Fig. \ref{performance_IRS_position} presents the system delay of different algorithms when the location of the IRS is moving along the $y$ axis. It is readily seen that when the IRS is far away from the users, the delay of our proposed algorithm approaches that of no IRS. When the IRS is located at the central location, the best latency performance is achieved.

Fig. \ref{performance_user_num} indicates the latency performance of various schemes versus the number of available users $J$. We can see that our proposed TTS algorithm achieves very close performance compared to the STS scheme and outperforms the others. We also observe that the gap between the proposed TTS scheme and max-to-max TTS scheme increases with the number of idle users. This is because the heuristic max-to-max matching strategy cannot find the globally optimal solution. Hence it suffers from performance degradation when the search space becomes larger.

\vspace{-1.0em}
\section{Conclusion}
In this letter, we investigated an IRS-aided D2D offloading system. We proposed a new two-timescale joint passive beamforming and resource allocation algorithm based on stochastic successive convex approximation to minimize the system latency while cutting down the heavy overhead for CSI feedback. Both the convergence property and the computational complexity of the proposed algorithm have been examined. Simulation results show that our proposed algorithm significantly outperforms the conventional benchmarks.
\begin{appendices}
	\section{Convergence Proof of Algorithm 1}
	Let $\bm{x}^\star(\boldsymbol{\theta},\mathbf{H})$ denote the optimal short-term variables under input $\boldsymbol{\theta}$ and $\mathbf{H}$.
	Then, the proof relies on the following lemma.
	\begin{lemma}
		We have
        \begin{equation}
         \|\bm{x}^\star(\boldsymbol{\theta}_1,\mathbf{H})-\bm{x}^\star(\boldsymbol{\theta}_2,\mathbf{H})\| \leq B_x\sqrt{\|\boldsymbol{\theta}_1-\boldsymbol{\theta}_2\|^2}, w.p.1, \label{continous_x}	
        \end{equation}
    \begin{equation}
    		\|f(\boldsymbol{\theta}_1,\bm{x}^\star(\boldsymbol{\theta_1},\mathbf{H}))-f(\boldsymbol{\theta}_2,\bm{x}^\star(\boldsymbol{\theta_2},\mathbf{H}))\| \leq B_f\sqrt{\|\boldsymbol{\theta}_1-\boldsymbol{\theta}_2\|^2},
    		\label{continous_f}		
    \end{equation}
    for any $\boldsymbol{\theta}_1,\boldsymbol{\theta}_2$ and some constant $B_x>0$, $B_f>0$, where w.p.1 is the abbreviation of with probability one.
	\end{lemma}

    \begin{IEEEproof}
    	There are totally $\frac{J!}{(J-I)!}$ kinds of situations for the user matching strategy and we define $\bm{x}^m(\boldsymbol{\theta},\mathbf{H}) \triangleq [\rho^\star_{i,j},u^m_{i,j}]$ to characterize one of the matching strategy. Then, we have
    	\begin{equation}
    		g(\boldsymbol{\theta},\bm{x}^{m_1}(\boldsymbol{\theta},\mathbf{H}))\neq g(\boldsymbol{\theta},\bm{x}^{m_2}(\boldsymbol{\theta},\mathbf{H})), \forall m_1\neq m_2, w.p.1. \label{probabilty0}
    	\end{equation}
        By considering that the probability density function of the joint distribution of $g(\boldsymbol{\theta},\bm{x}^{m_1}(\boldsymbol{\theta},\mathbf{H}))\times g(\boldsymbol{\theta},\bm{x}^{m_2}(\boldsymbol{\theta},\mathbf{H}))$ is bounded and the integral along the line $g(\boldsymbol{\theta},\bm{x}^{m_1}(\boldsymbol{\theta},\mathbf{H}))= g(\boldsymbol{\theta},\bm{x}^{m_2}(\boldsymbol{\theta},\mathbf{H}))\}$ equals zeros, \eqref{probabilty0} can be easily verified. We then prove that
        \begin{equation}
        \vspace{-1.0mm}
        \lim\limits_{\triangle\boldsymbol{\theta}\to\mathbf{0}} \|x^\star(\boldsymbol{\theta},\mathbf{H})-x^\star(\boldsymbol{\theta}+\triangle\boldsymbol{\theta},\mathbf{H})\| = 0,w.p.1.
        \end{equation}

        Based on \eqref{probabilty0}, we have $g(\boldsymbol{\theta},\bm{x}^{\star}(\boldsymbol{\theta},\mathbf{H})) < g(\boldsymbol{\theta},\bm{x}^n(\boldsymbol{\theta},\mathbf{H})), w.p.1, \forall \bm{x}^n(\boldsymbol{\theta},\mathbf{H}) \neq \bm{x}^{\star}(\boldsymbol{\theta},\mathbf{H})$. Moreover, since $g(\boldsymbol{\theta},\bm{x}^m(\boldsymbol{\theta},\mathbf{H})), \forall m$ are Lipschitz continuous functions with respect to $\boldsymbol{\theta}$. We have $\lim\limits_{\bigtriangleup\boldsymbol{\theta}\to\mathbf{0}} g(\boldsymbol{\theta}+\triangle\boldsymbol{\theta},\bm{x}^\star(\boldsymbol{\theta},\mathbf{H})) > \lim\limits_{\triangle\boldsymbol{\theta}\to\mathbf{0}} g(\boldsymbol{\theta}+\bigtriangleup\boldsymbol{\theta},\bm{x}^n(\boldsymbol{\theta},\mathbf{H})), w.p.1, \forall \bm{x}^n(\boldsymbol{\theta},\mathbf{H}) \neq \bm{x}^{\star}(\boldsymbol{\theta},\mathbf{H})$. As a result, we have \begin{equation}
        \vspace{-1.0mm} \lim\limits_{\bigtriangleup\boldsymbol{\theta}\to\mathbf{0}} x^\star(\boldsymbol{\theta}+\triangle\boldsymbol{\theta},\mathbf{H}) = x^\star(\boldsymbol{\theta},\mathbf{H}),w.p.1,
        \end{equation}
        thus \eqref{continous_x} is proved. Moreover, since $f(\boldsymbol{\theta},\bm{x}^\star(\boldsymbol{\theta},\mathbf{H}))$ can be written as $\mathbb{E}_{\mathbf{H}}\{\min_{m} g(\bm{\theta},\bm{x}^m(\boldsymbol{\theta},\mathbf{H}))\}$, \eqref{continous_f} holds immediately. Then, based on the Lemma 1 in \cite{stochastic_algorithm}, it can be verified that the proposed algorithm converges to a stationary point $w.p.1$.
    \end{IEEEproof}
\vspace{-1.0em}
\end{appendices}
\bibliographystyle{IEEEtran}
\bibliography{IEEEabrv,IRS_Offloading}

\end{document}